\documentstyle{article}

\renewcommand{\title}[1]{\newcommand{\IWANNTtitle}{#1}}
\renewcommand{\author}[1]{\newcommand{\IWANNTauthor}{#1}}
\newcommand{\affiliation}[1]{\newcommand{\IWANNTaffiliation}{#1}}
\renewcommand{\maketitle}{
\begin{center}
\mbox{ }\\[2.5 cm]
{\LARGE\bf \IWANNTtitle }\\[2 em]
{\normalsize\bf \IWANNTauthor} \\[1 em]
{\normalsize {\em \IWANNTaffiliation}}\\
\end{center}
\vspace{1 em}
}
\renewcommand{\thebibliography}[1]{
 \subsection*{References}
  \list
 {[\arabic{enumi}]}{\settowidth\labelwidth{[#1]}\leftmargin\labelwidth
 \advance\leftmargin\labelsep
 \usecounter{enumi}}
 \def\newblock{\hskip .11em plus .33em minus .07em}
 \sloppy\clubpenalty4000\widowpenalty4000
 \sfcode`\.=1000\relax
}

\begin{document}
\vspace*{-3.0cm}
\begin{center} \footnotesize Published in the Proceedings of the Workshop on
Nonlinear Evolution Equations and Dynamical Systems (NEEDS 93), Lecce, Italy,
{\it September 3-12, 1994.} \end{center}
\title{Nonlinear Dynamics in Distributed Systems}

\author{I. Adjali, J.L. Fern\'andez-Villaca\~nas and M. Gell}

\affiliation{Systems Research Division, BT Laboratories\\
Martlesham Heath, Ipswich IP5 7RE\\
 United Kingdom\\
{\tt iadjali@bt-sys.bt.co.uk}
}

\maketitle

\begin{abstract}
We build on a previous statistical model for distributed systems and formulate
it in a way that the deterministic and stochastic processes within the system
are clearly separable. We show how internal fluctuations can be analysed in a
systematic way using Van Kanpen's expansion method for Markov processes. We
present some results for both stationary and time-dependent states.
Our approach allows the effect of fluctuations to be explored, particularly in
finite systems where such processes assume increasing importance.
\end{abstract}

\section{Introduction}

With the increasing complexity of telecommunication and computational systems,
an urgent requirement is developing for theoretical frameworks for addressing
basic principles of distributed systems~\cite{Gel93}. At present there is
insufficient
understanding of principles required to predict performance, to explain
behaviour and to establish design methodologies~\cite{Kle86}. The substantial
vacuum in
theoretical bases for distributed communication and computational systems
stems largely from the historical preoccupation of computer and
telecommunication science with uniprocessor systems~\cite{Pow90}.

The emergence of large decentralised systems is giving rise to the need for a
general theoretic guide to the behaviour of large collections of locally
controlled, asynchronous and concurrent processes interacting with an
unpredictable environment. In particular this requires understanding the
relation between the overall behaviour of the distributed system and that of
its constituents, whose decisions are based upon local, imperfect, delayed and
conflicting information. In many other systems, particularly in nature and
societies, distributed systems with very complex behaviour and modes of
operation have evolved. There is a growing awareness that many of the
theoretical tools which have been developed with considerable success to
describe distributed systems in physics~\cite{Hak88}, particularly in condensed
matter physics, may be exploited in other fields, such as biology and
economics~\cite{Cal74}~\cite{Art88}~\cite{Bak91}~\cite{Ola93}.

\section{The Model}

A central feature of open systems is the non-linear nature of their dynamics,
which gives  rise to a rich repertoire of behavioural regimes ranging from
stable equilibrium to oscillations and chaotic states. One has to construct a
model which integrates both  the
deterministic evolution equation, responsible for the macroscopic behaviour of
the system, and the stochastic part  which deals with fluctuations within the
system.

A model was formulated for describing a self-organising
open computational system with resources, free agents and pay-off mediated
interactions~\cite{Adj93} that builds on as well as overcomes the limitations
inherent in previous work~\cite{Hub88}~\cite{Kep89}.

\begin{figure}[htb]
\vspace*{10cm}
\includegraphics{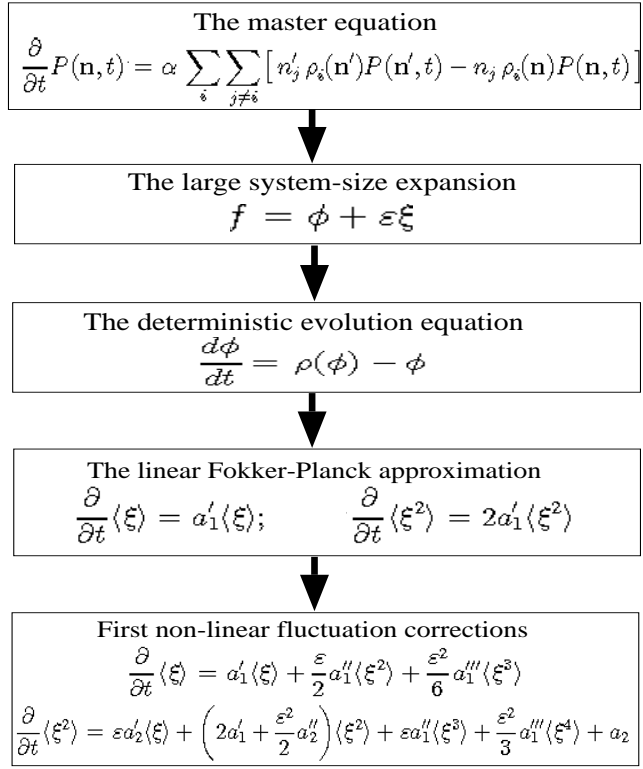}
\caption{Main steps in the analysis of the proposed model for an agent-resource
system}
\label{f1}
\end{figure}

Our approach allows the effects of fluctuations to be
investigated systematically in the form of a large-system size expansion due
to Van Kampen~\cite{Van61}~\cite{Van81}.
Figure~\ref{f1} shows an outline of the model; after writing down a
probabilistic evolution equation for a
general agent-resource system, we interpret
 the master equation obtained as describing a Markovian jump process and go on
to apply
Van Kampen's system size expansion. The deterministic equation for the
behaviour of the system arises
as the lowest-order term in the expansion and coincides
with the mean-field equation of Kephart {\it et al.}~\cite{Kep89}. The main
contribution of the fluctuations comes in the form of a linear Fokker-Planck
equation (FPE). Up  to this order the noise in the system is linear and the
solution of the master equation is given by a Gaussian distribution.
Non-linear effects of fluctuations are calculated as small perturbations to
the linear noise approximation. A detailed derivation of the equations in
Fig.~\ref{f1} can be found in Ref.~\cite{Adj93}.
In the following section some results for a two resource
system in both stationary and time-dependent states are presented and
discussed.
Section 4 summarises the main points and indicates directions for further work.

\section{Results and discussion}

Our main objective is to investigate
the approximation scheme for fluctuations based on the large system size
expansion of Van Kampen. The one-step Markovian
formulation of the problem allows us, in particular, to calculate the exact
probability distribution for time-independent solutions (see eq.~(12)
in~\cite{Adj93}). We
shall begin by restricting our numerical calculations to stationary solutions
of
the system in order to make a direct comparison with exact results and test
the validity of the approximation over a range of parameter values.

\begin{figure}[htb]
\vspace*{6cm}
\includegraphics{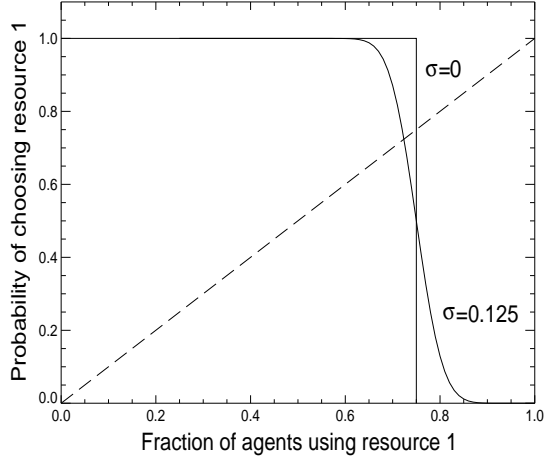}
\caption{The transition probability (from resource 2 to resource 1) $\rho(f)$
 corresponding to the pay-offs given in equ.~(1), for two values of the
uncertainty parameter $\sigma=\sigma_1=\sigma_2$. The intersection with the
line $\rho=f$ gives the solution
to the time-independent macroscopic equation.}
\label{f3}
\end{figure}

 The function $\rho$ represents the probability that an agent in the
system will find resource 1 to be more attractive than resource 2. In
general, the exact form of $\rho$ is not known and will depend on several
features
of the problem at hand, such as incomplete, uncertain or delayed information
about
the available resources. As a first example, we make $\rho$ a function of the
payoffs $G_{1}$ and
$G_{2}$ for using resources 1 and 2 respectively (as in~\cite{Kep89}),
\begin{equation}
G_1=7-f_1\;\;\;{\rm and}\;\;\;G_2=7-3f_2     \label{g1g2}
\end{equation}
These pay-off functions model a simple
competitive behaviour (opposing gradients) between agents so that the payoff
for using each
resource decreases with the number of agents already using the same resource.
 The system reaches a stability point when the two pay-offs are
equal so agents will prefer staying with the resource they are using. For $G_1$
and $G_2$  given in~(\ref{g1g2}) this optimal behaviour of the system occurs
for $f=0.75$,
that is, 75\% of all agents using resource 1. The decision region can be made
less sharply defined by introducing an uncertainty element in the payoff
evaluation. This can be achieved by introducing Gaussian noise with
standard deviation $\sigma$ around the true value of the pay-off.
If we assume that the agents' perception
of each resource is different, then there will be one uncertainty parameter for
each resource, $\sigma_1$ and $\sigma_2$. The resulting
transition probability $\rho$ is given by,
\begin{equation}
\rho=\frac{1}{2}\left[1+erf\left(\frac{G_1-G_2}{\sqrt{2}\sqrt{\sigma_1^2+\sigma_2^2}}\right)\right]  \label{rho2}
\end{equation}
\noindent and shown in Figure~\ref{f3} for $\sigma_1=\sigma_2=0.125$. The two
limiting cases of $\sigma_1=\sigma_2=0$ and
$\sigma_{1,2}=\infty$ correspond respectively to perfect knowledge ($f=0.75$)
and complete
lack of information on pay-offs, leading to the uniform distribution of agents
(\mbox{$f=0.5$}).
By approximating  $f$  with its deterministic contribution $\phi$,
we obtain a graphical solution of the deterministic equation
($\rho(\phi)=\phi$) represented
in Figure~\ref{f3} by the crossing point between the curves $\rho(\phi)$ and
$\phi$. This point
gives the equilibrium solution which now, due to a non-zero value of the
uncertainty, is
slightly offset from the optimal value $f=0.75$. The macroscopic value of $f$
for $\sigma_1=\sigma_2=0.125$ is $\phi=0.724$.

\begin{figure}[htb]
\vspace*{7.0cm}
\includegraphics{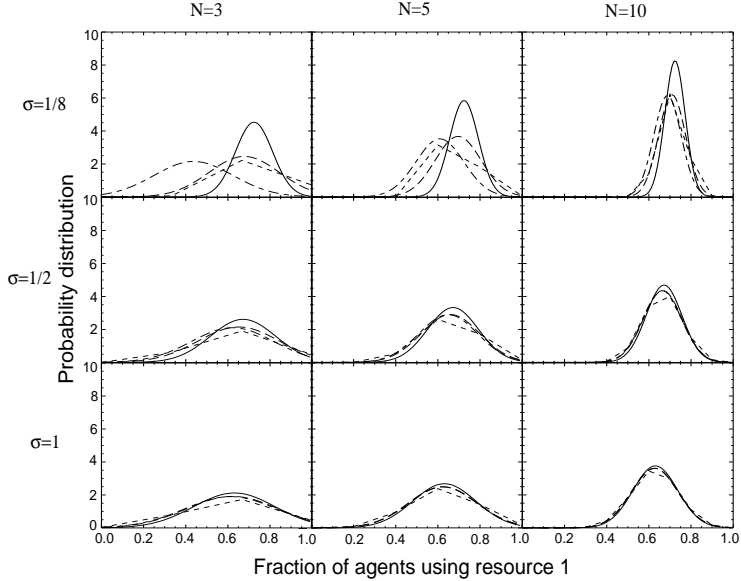}
\caption{The time-independent probability distribution for three different
values of N and three different values of $\sigma=\sigma_1=\sigma_2$, in three
orders of the large
system-size expansion; full line: mean-field (including linear noise) result,
long dash: first order non-linear corrections included, long dash-short dash:
second order non-linear corrections included, short dash: exact solution.}
\label{f4}
\end{figure}

In order to see how the Van Kampen approximation depends on the uncertainty
parameters,  we have plotted the time-independent probability distributions
for different orders in the approximation as well as the exact distribution,
for three values of $\sigma=\sigma_1=\sigma_2$ (vertically) and three values of
the number of agents $N$ (horizontally) in Figure~\ref{f4}.
We can draw the following conclusions: The approximation works
reasonably well  for all values of $\sigma$ considered; the first  order
non-linear corrections are sufficient for  correctly estimating fluctuation
effects in the system, especially if the uncertainty parameter is not too
small. Furthermore it seems that the approximation is best suited for systems
with a moderate value of the uncertainty parameter $\sigma(\approx~0.5)$, where
non-linear effects of fluctuations, although significant, converge rapidly in
the expansion. This may be the range of $\sigma$ to look for in realistic
systems,
where agents are neither expected to have perfect knowledge nor be completely
ignorant about  the pay-offs of their transactions.

\begin{figure}[htb]
\vspace*{8.5cm}
\includegraphics{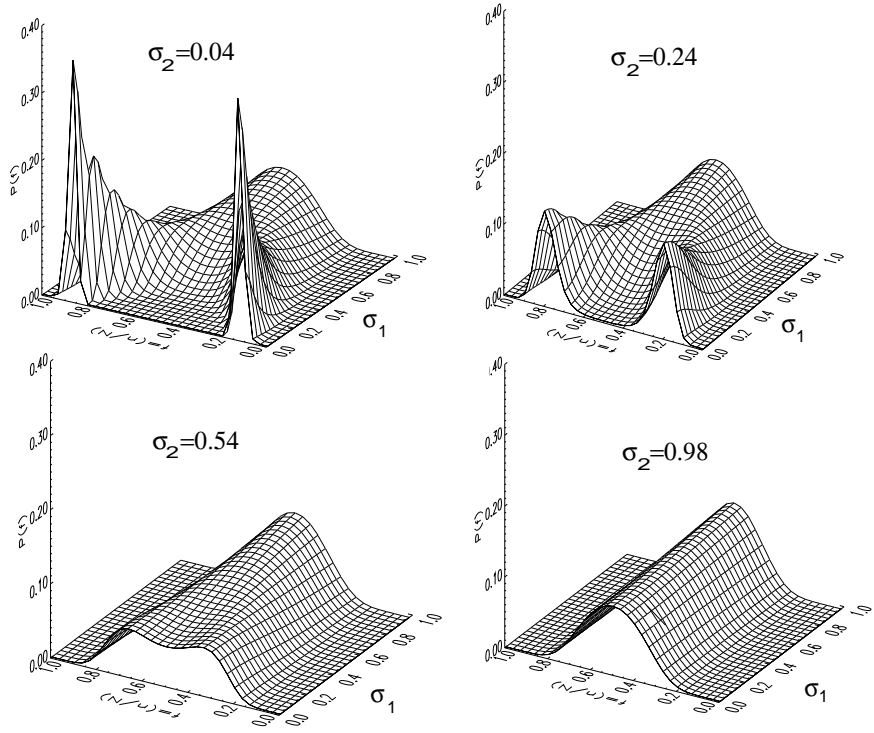}
\caption{Effect of the uncertainty parameters $\sigma_1$ and $\sigma_2$ on a
bistable system.}
\label{f6}
\end{figure}

So far in our analysis we have only looked at systems with a single
macroscopic stable behaviour, a consequence of the unique (stable) fixed point
occuring at the intersection between the linear pay-off functions $G_1$ and
$G_2$. This simple competitive behaviour can be changed by
making the pay-off functions non-linear, ie. introducing cooperation as well
as competition between agents in the system. Whereas competition meant that
agents would favour a resource if it had less agents using it, cooperation is
expressed by an increased pay-off when a resource is used by more agents. The
interplay of these two tendencies through non-linear pay-offs leads to a
richer range of possible behaviours in the system. We treat here the example
of a bistable system (arising from cubic pay-offs) (see Figure~\ref{f6}).
Depending on which side of the mid-point the initial distribution is,
the system will eventually settle in one of the macroscopic states
characterised by the two peaks in the time-independent probability
distribution.

The system's dynamics depends notably on the different values of the
uncertainty parameters $\sigma_1$ and $\sigma_2$. In Figure~\ref{f6} we have
shown
the dependence on $\sigma_1$ for four different values of $\sigma_2$, using the
non-linear (cubic) pay-offs.
By increasing $\sigma_1$, $\sigma_2$ or both, the two peaks are seen to
gradually
get closer to each other and merge into a single (symmetric) peak.
These critical values of $\sigma_1$ and $\sigma_2$ can be found by inspection
of the
time-independent macroscopic equation; they play the role of  control
parameters which can change qualitatively the dynamical phase space of the
system, in this case from a system with two attractors to a system with a
single one. This is reminiscent to phase transitions in physical systems such
as the spontaneous magnetisation of a ferromagnetic system which happens by
lowering the temperature below a critical value (Curie temperature). Above
this value the overall magnetisation is zero and symmetric while below it
there are two possible states of opposite magnetisation. By choosing one state
or the other, the system breaks its spatial symmetry, just like by decreasing
$\sigma_1$ or $\sigma_2$ below their critical values in the agent-resource
system we see a sudden
transition from an equal distribution of agents on the two resources to a
definite bias towards one or the other.

\subsubsection*{Time-dependent solution}
So far we have described results derived from the time-independent
simplification
of the model~\cite{Adj93}. In order to extend our simulations to account for
time-dependent
behaviour we began by studying the evolution of a system with two resources and
a number of agents
$10<N<50$. We used the deterministic equation (see Figure~\ref{f1}) with the
linear pay-offs as in~(\ref{g1g2}).
Each pay-off is described by a linear equation in the percentage of occupancy
$f$ of resource 1
(bear in mind that the system is closed, thus $f_2=1-f$) as,
\begin{equation}
G_1=af+b\;\;\;{\rm and}\;\;\;G_2=cf+d\label{lin1}
\end{equation}
\noindent If we start from an intial distribution
$f$, which will depend on $\sigma_1$ and $\sigma_2$, the less-dominant resource
($G_2$) mutates (increases)
its slope and intercept proportionally to $\Delta f=f-0.5$. In order to
constrain the system we postulate that these
increases are equally matched by the decreases in slope and intercept for
resource~1,
\begin{equation}
G_2=(c+\Delta c)f+(d+\Delta d)\;\;\;{\rm and}\;\;\;G_1=(a-\Delta c)f+(b-\Delta
d) \label{lin2}
\end{equation}
\noindent where,
\begin{equation}
\Delta c= \gamma \Delta f+\delta\;\;\;{\rm and}\;\;\;\Delta d=\alpha \Delta
f+\beta  \label{delt}
\end{equation}
\noindent each has two contributions: one depending on how badly they are
loosing to the competing
resource and another random component introducing noise. For simplicity we have
adopted $\alpha=\gamma$
and $\delta=\beta$.

The competing process goes as follows: the deterministic equation gives an
initial distribution $f$
that allows each resource to calculate how much they have to mutate their
pay-offs to become more
attractive to the agents. The new pay-offs are re-introduced, together with
$\sigma_1$ and $\sigma_2$,
to calculate the new probability $\rho$ which is used to solve the
deterministic
and the fluctuations equations simultaneously.

As in biology, the rate at which mutations happen is fundamental to achieving
an evolutionary
improvement. In this simple case we have observed that after a mutation, and
whichever initial configuration we start from,
equilibrium is always reached after 2 units of time. This is of the order of
the relaxation time of the evolution equations.
\begin{figure}[htb]
\vspace*{6cm}
\includegraphics{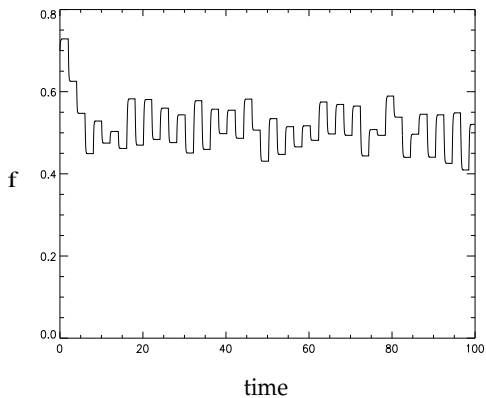}
\caption{Evolution in a competitive system with linear pay-offs}
\label{f7}
\end{figure}
In Figure~\ref{f7} we show the evolution of a system with two resources with
pay-offs as in ~(\ref{g1g2}),
with $\sigma_1=\sigma_2\approx 0$ and a relaxation of 2 units of time. The
simulation starts from
resource~1 having its maximum of probability centered on 73\% of the number of
agents. At $t=5$ resource~2
has pulled back and is winning more than half of the agents, while at $t=10$
the situation goes
back to a balance (50\%). After this point the simulation is dominated by the
noise introduced by the
$\beta$ and $\delta$ values (see~(\ref{delt})).

\section{Conclusions}

We have studied a model for market-like agent-resource systems,
whose formulation is based on one-step Markov processes and the large
system-size expansion of the master equation due to Van
Kampen~\cite{Van61}~\cite{Van81}. Our
formulation enables a systematic treatment of fluctuations to be carried out.
A deterministic equation governing the dynamics of the system arises as the
lowest order contribution in the
expansion, and coincides with the equation obtained in the mean-field approach
{}~\cite{Kep89}. The next order term gives the main contribution of the
fluctuations in the form of a linear Fokker-Planck equation. The probability
distribution
describing the dynamics of the system is therefore a Gaussian
distribution to this order in the expansion. Higher order terms are included to
provide non-linear corrections to the FPE.
the lowest order non-linear corrections
represent fluctuations due to individual agents in the system.
Higher order corrections are crucial when the number of agents is relatively
small and mean-field theory inadequate.

To test the approximation in the case of our agent-resource system, we have
taken a system with two resources and considered time-independent states.
Taking numerical
values as in ref.~\cite{Kep89} shows full agreement between our exact
theoretical
results and the corresponding Monte Carlo simulations performed in
ref.~\cite{Kep89}.
Sensitivity to
accuracy of the information available to agents was also studied and the main
observation is that higher uncertainty leads to the suppressing of non-linear
noise effects.
In view
of the results obtained we conclude that the approximation works generally
well and can therefore be reliably used for time-dependent solutions.

We have modelled
time evolution by making two resources with linear pay-offs compete for the
agents; after
an initial period of instability the system adopts a {\it tit-for-tat} cycle
where
one resource dominates the other only to give way to the competing one after a
fixed relaxation time interval.
The time-dependent solutions for a small number of agents (for which
first and higher order approximations are required) have been
studied~\cite{Fer93}. In that study, more realistic descriptions of the
pay-offs in terms of systems' measurable properties and more
sophisticated evolution mechanisms are described.

We should note, however, that Van Kampen's approximation scheme is not suited
for the treatment of fluctuations in situations involving instabilities or
critical behaviour. In other words, the system size expansion is valid only
when there is one globally stable macroscopic solution (such as a simple
competitive system) or in the immediate vicinity of a
locally stable solution. In general, however, a system with
multiple minima requires a different treatment of fluctuations near instability
points. This important issue will be addressed in a separate work.

\end{document}